\documentclass[a4paper,11pt]{article}
\usepackage{graphicx,epsfig}
\usepackage{grffile}
\usepackage{amsmath,amssymb}
\usepackage{array}
\usepackage{color}
\usepackage{slashed}
\usepackage{enumitem}
\usepackage[utf8]{inputenc}
\usepackage[normalem]{ulem}
\usepackage{cite}

\newcommand{\ord}[1]{\mathcal{O}\left({#1}\right)}

\title{\Large\bf\boldmath Majorana vs Pseudo-Dirac Neutrinos at the ILC}

\date{}
\author{P.~Hern\'andez$^{a}$, J.~Jones-P\'erez$^{b}$, O.~Suarez-Navarro$^{b}$
\\[0.5 cm]
$^a${\em\normalsize Instituto de F\'isica Corpuscular (IFIC), CSIC-Universitat de Val\`encia,} \\
{\em\normalsize Apartado de Correos 22085, E-46071 Valencia, Spain} \\
$^b${\em\normalsize Secci\'on F\'isica, Departamento de Ciencias, Pontificia Universidad Cat\'olica del Per\'u,} \\
{\em\normalsize  Apartado 1761, Lima, Peru}
} 

\begin{document}

\maketitle

\begin{abstract}
\noindent
Neutrino masses could originate in seesaw models testable at colliders, with light mediators and an approximate lepton number symmetry. The minimal model of this type contains two quasi-degenerate Majorana fermions forming a pseudo-Dirac pair. An important question is to what extent future colliders will have sensitivity to the splitting between the Majorana components, since this quantity signals the breaking of lepton number and is connected to the light neutrino masses. We consider the production of these neutral heavy leptons at the ILC, where their displaced decays provide a golden signal: a forward-backward charge asymmetry, which depends crucially on the mass splitting between the two Majorana components. We show that this observable can constrain the mass splitting to values much lower than current bounds from neutrinoless double beta decay and {\it natural} loop corrections.
\end{abstract} 

\section{Introduction}

Extensions of the Standard Model that can explain neutrino masses  \cite{Minkowski:1977sc,GellMann:1980vs,Yanagida:1979as,Mohapatra:1979ia} are well motivated leads to the new physics realm. The mass scale of the neutrino mass mediators is unknown, and the possibility that they could be light enough to be produced and tested in laboratory experiments has been extensively discussed in the literature. Particularly interesting signals of this type of new physics are displaced vertices~\cite{Helo:2013esa,Blondel:2014bra,Cui:2014twa,Gago:2015vma,Duarte:2016caz,Antusch:2016vyf,Caputo:2017pit,Antusch:2017hhu,Abada:2018sfh}, since usually such light mediators are also very weakly coupled and have long lifetimes.

Present and future colliders have the opportunity to discover or constrain interesting regions of parameter space in these models~\cite{Deppisch:2015qwa}, particularly in connection with low-scale leptogenesis \cite{Akhmedov:1998qx,Asaka:2005pn,Shaposhnikov:2008pf,Canetti:2012zc,Canetti:2012kh,Asaka:2011wq,Shuve:2014zua,Abada:2015rta,Hernandez:2015wna,Hernandez:2016kel,Drewes:2016gmt,Drewes:2016jae,Hambye:2016sby,Ghiglieri:2017gjz,Asaka:2017rdj,Hambye:2017elz,Abada:2017ieq,Ghiglieri:2017csp}. The possibility to test leptogenesis scenarios is however very challenging due to the large parameter space that can affect the generated  baryon asymmetry. The putative discovery of the neutrino mass mediators and the measurement of their properties will be essential to achieve this goal. Particularly important questions are establishing the Majorana nature of the heavy neutral leptons expected in the Type I seesaw model and measuring their mass spectrum~\cite{Anamiati:2016uxp} and flavour mixings~\cite{Caputo:2017pit}.

Determining the Majorana nature for on-shell particles is in principle straightforward, it is sufficient to observe their lepton number violating decays (LNV). However, light neutrino mass mediators with sufficiently large mixings  require an approximate lepton number symmetry to avoid fine-tuning \cite{Wyler:1982dd,Kersten:2007vk,Gavela:2009cd,Ibarra:2010xw,LopezPavon:2012zg,Moffat:2017feq}. This implies that mediators come in quasi-degenerate pairs (i.e.\ pseudo-Dirac particles) that interfere destructively to cancel the LNV decays~\cite{Kersten:2007vk}. We expect on general grounds that the  cancellation of LNV decays will be effective provided the mass splitting, $\delta M$, of the pseudo-Dirac pair is small compared to their decay width, $\Gamma$, which typically requires a strong degeneracy. On the other hand, if LNV decays are not suppressed, i.e.\ $\delta M \gg \Gamma$, it is interesting to understand to what extent  their existence can be established. 

In this paper we consider the production of pseudo-Dirac neutrinos $N$  in a $e^+ e^-$ collider, in processes such as $e^+ e^- \rightarrow N \nu$, leading to a displaced semileptonic decay, $N \rightarrow l^\pm j j$.  The total number of positive and negative leptons in the displaced vertex is not a good discriminator of the Majorana nature, because the final light neutrino or antineutrino goes undetected. However, the  pseudorapidity distribution of the final lepton changes drastically in the two cases~\cite{delAguila:2005pin} and therefore is a good discriminator. Other recent proposals to use angular information to test the Majorana nature have been recently discussed in \cite{Arbelaez:2017zqq,Balantekin:2018ukw}. 

The paper is organized as follows. In Section~\ref{sec:model} we review the minimal Type I seesaw model, where we define our notation and link the heavy neutrino mass splitting $\delta M$ with LNV parameters. In Section~\ref{sec:constraints}, we review the current most stringent constraints on $\delta M$ that come from neutrinoless double beta decay, and the requirement of no fine-tuning between tree and loop corrections to the light neutrino masses. 
In Section~\ref{sec:hdecay} we examine the process $e^+e^-\to \nu N\to\nu l^\pm W^{\mp *}$, and 
show that the LNV contribution effectively vanishes when the mass splitting (and width difference) goes to zero.
In Section~\ref{sec:ILC} we study this process at the ILC, and quantify the forward-backward charge asymmetry of the lepton as a function of $\delta M$. Backgrounds are avoided by requiring the heavy neutrino signature to include a displaced vertex. The putative observation of such an asymmetry implies strong bounds on $\delta M$ of the order of decay width, that could useful to constrain resonant leptogenesis scenarios \cite{Pilaftsis:2005rv}.

\section{The Minimal Type I Seesaw Model}
\label{sec:model}

The minimal way of generating neutrino masses is achieved by extending the SM with two heavy Majorana spinors, singlets under the gauge symmetries, which are usually identified as sterile neutrinos or neutral heavy leptons. By imposing a lepton number symmetry \cite{Wyler:1982dd,Mohapatra:1986bd}, one can assign the two Majorana fields opposite lepton number charges, such that the Lagrangian reads:
   \begin{eqnarray}
{\mathcal L} = {\mathcal L}_{SM}- \sum_{\alpha=e,\mu,\tau} \bar L^\alpha Y_{\alpha 1} \tilde\Phi N_{1R} -  \frac{1}{2}\overline{N}^{c}_{1R} M N_{2 R}+ h.c. \nonumber
\label{eq:lag}
\end{eqnarray}
The two degenerate Majorana spinors can be combined into one massive Dirac neutrino. After electroweak symmetry breaking, the allowed terms in the Lagrangian lead to the following mass matrix in the basis $(\nu_\alpha, N_{1 R}, N_{2 R})$:
\begin{equation}
\label{eq:Diracmass}
 M_\nu=\left(\begin{array}{ccc}
0 & m_D & 0 \\ 
m_D^T & 0 & M \\
0 & M & 0
\end{array}\right),
\end{equation}
where $(m_D)_\alpha = \tfrac{v}{\sqrt2}Y_{\alpha 1}$ is a three component vector. Even though lepton flavour is violated in this limit, lepton numbers is not and the SM neutrinos remain massless. 

If the above structure is perturbed by slightly breaking the lepton number symmetry, the heavy Majorana pair are no longer degenerate and the Dirac fermion becomes a pseudo-Dirac one. The perturbed mass matrix can be written:
\begin{equation}
\label{eq:massmat}
 M_\nu=\left(\begin{array}{ccc}
0 & m_D & \varepsilon \\ 
m_D^T & \mu' & M \\
\varepsilon^T\,  & M & \mu
\end{array}\right)
\end{equation}
where $\mu$, $\mu'$ and $\varepsilon$ are lepton number violating (LNV) terms, which can be kept small in a technically-natural way. In particular, $\mu$ and $\mu'$ contribute to the splitting of the heavy states. When all the LNV parameters are small $ \mu, \mu', \varepsilon\ll M$, we expect LNV processes to be suppressed accordingly \cite{Kersten:2007vk}.

These textures are well known. Setting $\mu=\mu'=0$ leads to the Linear Seesaw~\cite{Malinsky:2005bi}, setting $\varepsilon=\mu'=0$ corresponds to the Inverse Seesaw~\cite{Wyler:1982dd,Mohapatra:1986bd}, while $\varepsilon=0$ is sometimes called Extended Seesaw~\cite{Kang:2006sn,LopezPavon:2012zg}. A radiative model with $\varepsilon=\mu=0$ has also been discussed in~\cite{Dev:2012sg}. Any of the three terms imply the existence of LNV processes, and, as long as $\varepsilon\neq 0$, one can explain the two measured light neutrino mass differences~\cite{Gavela:2009cd}.

In contrast with the standard seesaw, the smallness of neutrino masses does not require a large hierarchy $m_D \ll M$, it is sufficient to suppress two LNV parameters: $\mu, \varepsilon\ll M$. On the other hand, the splitting of the heavy states also depends on $\mu'$, which has a negligible impact on light neutrino masses at tree level. If $\mu'$ is large, one can have a sizeable splitting of the heavy states and can expect unsuppressed contributions to LNV processes. Nevertheless, in this case large loop corrections to light neutrino masses are also expected, which must be kept under control~\cite{LopezPavon:2012zg}.



Even though the parametrization in Eq.~(\ref{eq:massmat}) is useful in order to understand the role of LNV terms, one can also use  a more convenient one involving the physical neutrino masses and the mixing angles~\cite{Casas:2001sr}. The extension proposed in~\cite{Donini:2012tt} is the one we shall use in this work. Here, the neutrino mixing matrix is divided into four blocks, which can be written\footnote{This work uses expressions valid for a normal ordering of the SM neutrino masses, for equivalent expressions with inverted ordering, one can see~\cite{Donini:2012tt,Gago:2015vma}.} in all generality as:
\begin{align}
 \label{eq:mixingmats}
 U_{a\ell} &= U_{\rm PMNS}\left(\begin{array}{cc}
 1 & 0 \\
 0 & H
\end{array}
\right)~, &
 U_{ah} &= i\,U_{\rm PMNS}\left(\begin{array}{c}
 0 \\
 H\,m_\ell^{1/2}R^\dagger M_h^{-1/2}
\end{array}\right)~, \nonumber\\
 U_{s\ell} &= i\left(\begin{array}{cc}
 0 & \bar H M_h^{-1/2}\,R\,m_\ell^{1/2}
\end{array}\right)~, &
 U_{sh} &= \bar H~.
\end{align}
Here, the labels $a=(e,\,\mu,\,\tau)$ and $s=(s_1,\,s_2)$ on the mixing matrices refer to the active (SM) and sterile neutrino interaction states, respectively, while $\ell=(1,\,2,\,3)$ and $h=(4,\,5)$ refer to the light and heavy neutrino mass eigenstates. $U_{\rm PMNS}$ is a unitary matrix which represents the light neutrino mixing, up to corrections from non-unitarity. As we are including two sterile neutrino states, only two light neutrinos acquire mass. This information is encoded in the $2\times2$, diagonal $m_\ell$ and $M_h$ matrices, which contain the light and heavy neutrino masses, respectively. The $R$ matrix, originally introduced in~\cite{Casas:2001sr}, is in this case a 2$\times$2 orthogonal complex matrix, fully defined in terms of the complex angle: $\theta_{45}+ i \gamma_{45}$.  Finally the $H, \bar{H}$ hermitian matrices are defined by
\begin{eqnarray}
\label{eq:hreal}
H &=& \left(I+m_\ell^{1/2}\,R^\dagger\,M_h^{-1}\,R\,m_\ell^{1/2}\right)^{-1/2}, \nonumber\\
{\bar H} &=& \left(I+M_h^{-1/2}\,R\,m_\ell\,R^\dagger\,M_h^{-1/2}\right)^{-1/2}~,
\end{eqnarray}
and contain the violations of unitarity. As we will show, large $\gamma_{45}$  is in one-to-one correspondence to the  
$\mu, \varepsilon \ll M \sim m_D$ region.  It is this corner of parameter space where the  flavour mixings of the heavy states can be larger than what the naive seesaw
scaling would suggest, i.e.\ $|U_{ah}|^2 \sim m_l/M_h$, and therefore offers better detection prospects. 

With large $\gamma_{45}$, present constraints  imply a strong upper bound on $x\equiv (\sqrt{m_{2,\,3}/M_{4,\,5}}\cosh\gamma_{45})$. Assuming $H\sim I$, an expansion in $x$ gives:
\begin{eqnarray}
\label{eq:mixapprox}
U_{a4} &\simeq &  \pm Z^{\rm NH}_a\sqrt{\frac{m_3}{M_4}}\cosh\gamma_{45}\,e^{\mp i\theta_{45}} +{\mathcal O}(x^3)~, \nonumber\\
U_{a5} &\simeq &  i\,Z^{\rm NH}_a\sqrt{\frac{m_3}{M_5}}\cosh\gamma_{45}\,e^{\mp i\theta_{45}}+{\mathcal O}(x^3)~, \nonumber
\end{eqnarray}
with:
\begin{equation}
 Z^{\rm NH}_a \equiv (U_{\rm PMNS})_{a 3}\pm i\sqrt{\frac{m_2}{m_3}}(U_{\rm PMNS})_{a 2}. 
\end{equation}
One finds that for heavy neutrino masses of the order of the GeV, $\gamma_{45}$ is bounded to values lower than 10 by LFV experiments, such as $\mu\to e\gamma$ and $\mu-e$ conversion~\cite{Gago:2015vma}. $\theta_{45}$ on the other hand remains unconstrained.

The full neutrino mass matrix, in the large $\gamma_{45}$ limit, has the general form:
\begin{equation}
\label{eq:massmatparam}
 M'_\nu=\left(\begin{array}{ccc}
0 & (m_D')_{a4} & (m_D')_{a5} \\ 
(m_D')^T_{a4} & M_4 & {\mathcal O}(x^2) \\
(m_D')^T_{a5} & {\mathcal O}(x^2) & M_5
\end{array}\right)~,
\end{equation}
where again $(m_D')_{ah}$ are three component vectors: 
\begin{eqnarray}
\label{eq:mDnew}
 (m'_D)_{a 4} &\simeq& \pm (Z^{\rm NH}_a)^*\sqrt{m_3 M_4}\cosh\gamma_{45}\,e^{\mp i\theta_{45}}+{\mathcal O}(x^3)~, \\
 (m'_D)_{a 5} &\simeq& -i (Z^{\rm NH}_a)^*\sqrt{m_3 M_5}\cosh\gamma_{45}\,e^{\mp i\theta_{45}}+{\mathcal O}(x^3)~.
\end{eqnarray}
With this, a field redefinition can readily put the neutrino mass matrix in the form of Eq.~(\ref{eq:massmat}). For example, if $\gamma_{45}>0$, we find $V^T M'_\nu \,V=M_\nu$, with:
\begin{equation}
 V=\left(\begin{array}{ccc}
I & 0 & 0 \\
0 & -i\cos\theta & i\sin\theta \\
0 & \sin\theta & \cos\theta
\end{array}
\right)
\end{equation}
and $\tan\theta=\sqrt{M_5/M_4}$. In this case, we find $\mu'=\delta M\equiv M_5-M_4$, while $\mu$ and $\varepsilon$ become of the order of the neglected terms, and therefore very small\footnote{Evidently, if $\mu'$ is very small, one cannot neglect $\mu$ and $\varepsilon$ in the calculation of $\delta M$. This would lead to a lower bound on the mass splitting, since the latter parameters are fixed by light neutrino masses.}. Thus, as was shown in~\cite{Lopez-Pavon:2015cga}, we find that $\mu'$ can encode a large violation of lepton number without affecting significantly the light neutrino masses at tree level, and that this effect depends on the mass splitting between the heavy neutrinos.

\section{Current Constraints on Mass Splittings}
\label{sec:constraints}

The mass splitting of the heavy neutrinos is connected primarily to $\mu'$ in the region of interest and LNV processes such as neutrinoless double beta decay ($0\nu\beta\beta$) are very sensitive to it~\cite{Mitra:2011qr, LopezPavon:2012zg}. Thus, the non-observation of this process can set an upper bound on the splitting. In addition, loop corrections to the light neutrino masses have also been shown to be sensitive to $\mu'$~\cite{ LopezPavon:2012zg,Lopez-Pavon:2015cga}, so requiring no unnatural cancellation between tree level and one loop corrections to neutrino masses can also severely constrain the value of the splitting. 

Provided that loop corrections to neutrino masses can be neglected, the total contribution to $0\nu\beta\beta$ can be written as~\cite{Blennow:2010th,Gago:2015vma}:
\begin{eqnarray}
\label{eq:neutrinoless.basic}
A_{\beta\beta}&\propto& \sum_{i=1}^3{m_{i}\,U_{ei}^2\,\mathcal{M}^{0\nu\beta\beta}(0)} + \sum_{i=4}^5{M_i\,U_{e i}^2\,\mathcal{M}^{0\nu\beta\beta}(M_i)} \nonumber \\
&\propto& m_{\beta\beta}\,\Delta\mathcal{M}(0,\,M_5)
+M_4\,U_{e4}^2\,\Delta\mathcal{M}(M_4,\,M_5) ~,
\end{eqnarray}
where $m_{\beta\beta}$ is the light neutrino contribution, $m_{\beta\beta}=\sum_{i=1}^3m_{ i}\,U_{ei}^2$, and we have used the exact relation:
\begin{equation}
\sum_{i=1}^3m_{ i}\,U_{ei}^2+ \sum_{i=4,5} M_i\, U_{ei}^2 = 0
\end{equation}
which follows from structure of the full neutrino mass matrix, Eq.~(\ref{eq:massmat}), that has vanishing $ee$ element as a result of gauge invariance. The nuclear matrix element (NME), $\mathcal{M}^{0\nu\beta\beta}$, is a function of the mass of the virtual neutrino. Finally, we have defined $\Delta\mathcal{M}(M_a,\,M_b)\equiv \mathcal{M}^{0\nu\beta\beta}(M_a)-\mathcal{M}^{0\nu\beta\beta}(M_b)$.

The NMEs remain practically constant for neutrino masses much lower than 100 MeV, which is the typical momentum transfer of this process. For larger masses the NMEs decrease with the inverse of $M_i^2$~\cite{Blennow:2010th}. Important limits are:
\begin{equation}
\begin{array}{lllll}
\Delta\mathcal{M}(M_a,M_b)&\simeq& \mathcal{M}^{0\nu\beta\beta}(0),  & & M_a\ll100~{\rm MeV} \ll M_b, \\
&\simeq& 0,   & & M_{a,b} \ll100~{\rm MeV},\\
& = & 0,&& M_a =M_b.
\end{array}
\end{equation}
If both neutrinos are much lighter than 100~MeV, the amplitude of this process vanishes, as results from Eq.~(\ref{eq:neutrinoless.basic}). 
If both neutrino masses are larger than 100~MeV, the heavy neutrino contribution is suppressed by the NME, but the second term  in Eq.~(\ref{eq:neutrinoless.basic}) can be sizeable for masses up to $\ord{10\,{\rm GeV}}$, depending on the value of $U^2_{e4}$.  This term vanishes however if $M_4$ and $M_5$ are degenerate. As we have seen, this essentially means setting $\mu'\to0$ and suppressed LNV. 

\begin{figure}[tbp]
\centering
\includegraphics[width=0.6\textwidth]{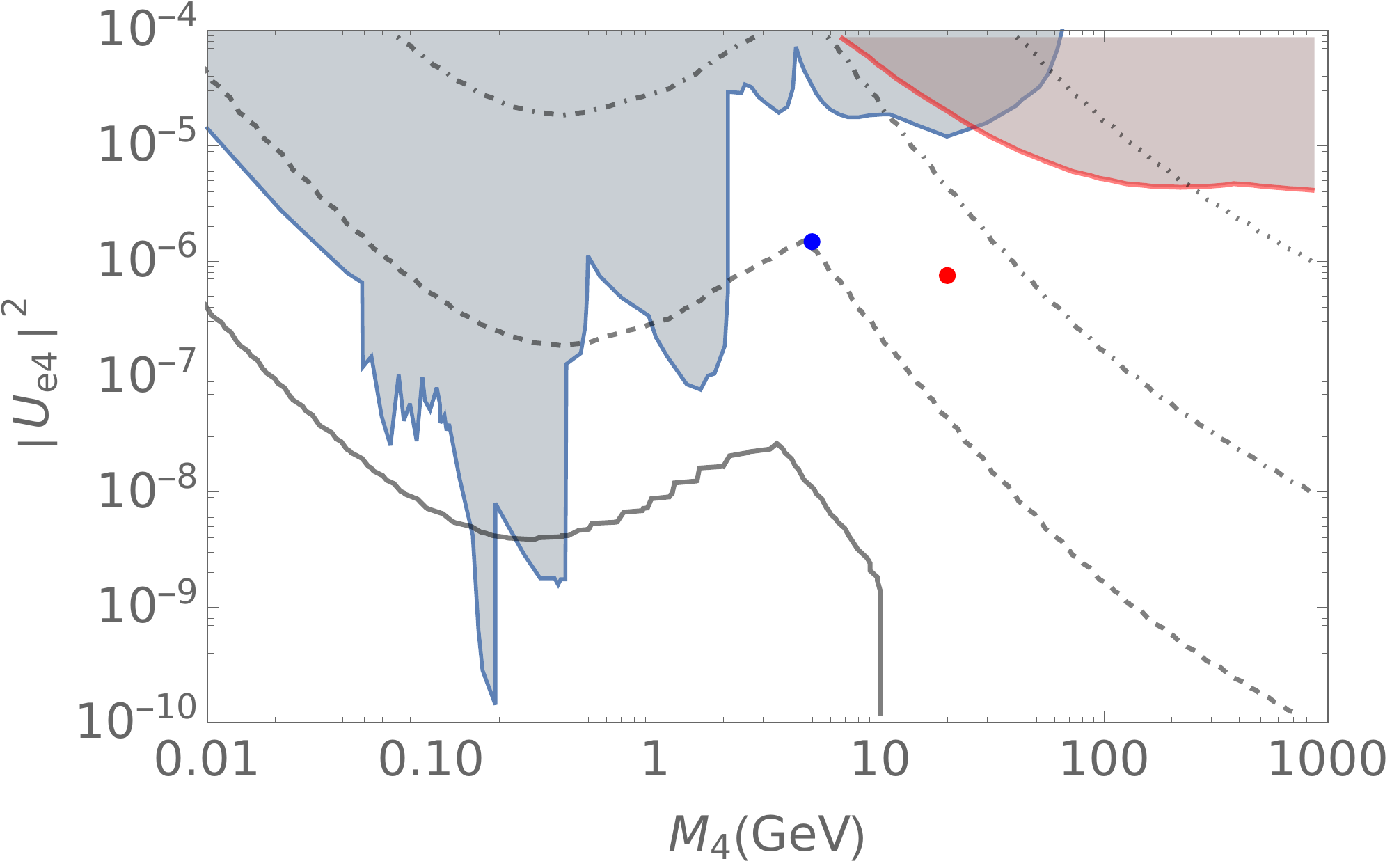}
\caption{Contours  of constant ${\rm Max}(\delta M/M_4)$, the maximum possible mass splitting allowed by loop and $0\nu\beta\beta$ constraints, on the plane $(M_4,|U_{e4}|^2)$. We show contours for ${\rm Max}(\delta M/M_4)=1,\,10^{-2},\,10^{-4},\,10^{-6}$ in solid, dashed, dash-dotted and dotted lines, respectively. Any point above a given contour would require more degeneracy than that corresponding to the contour label. The shaded blue region is ruled out by direct searches, taken from~\cite{Abada:2017jjx}, while the red region is ruled out by LFV experiments~\cite{TheMEG:2016wtm,Bertl:2006up}. The blue and red dots correspond to the \textit{light} and \textit{heavy} benchmarks, respectively, defined in Section~\ref{sec:ILC}.} 
\label{fig:combi}
\end{figure}
On the other hand, loop corrections to light neutrino masses have also been extensively studied in the past~\cite{Pilaftsis:1991ug,Grimus:2002nk,AristizabalSierra:2011mn,Lopez-Pavon:2015cga}. In our approximation these are given by:
\begin{equation}
 \delta m_{\rm loop}=\frac{g^2}{64\pi^2m_W^2}m'_D\,M_h^{-1}
 \left(m_H^2\ln\left[\frac{M_h^2}{m^2_H}\right]+3m_Z^2\ln\left[\frac{M_h^2}{m_Z^2}\right]\right)m^{\prime\, T}_D
\end{equation}
where $m_Z$ and $m_{H}$ are the $Z$ and Higgs boson masses, respectively. In this case, due to the structure of $m'_D$ in Eq.~(\ref{eq:mDnew}), the leading term again vanishes if the mass splitting goes to zero. Conversely requiring the 
loop corrections to be sufficiently small, gives an upper bound on the heavy neutrino mass splitting or on the heavy neutrino mixing for a fixed value of the splitting. 

In Figure~\ref{fig:combi} we show iso-contours of maximum splitting ${\rm Max}(\delta M/M_4)$ on the plane $(M_4, |U_{e4}|^2)$, as can be derived from current constraints on $0\nu\beta\beta$~\cite{KamLAND-Zen:2016pfg} and loop corrections. To obtain the limits from $0\nu\beta\beta$, we neglect the light neutrino contribution, since it is significantly below the present experimental limit in this model, in other words, we only consider the second term of Eq.~(\ref{eq:neutrinoless.basic}). For each point, we require $A_{\beta\beta}/\mathcal{M}^{0\nu\beta\beta}(0)<165~$meV \cite{KamLAND-Zen:2016pfg} , which becomes the most important constraint  for $M_4\lesssim1$~GeV. Moreover, we require loop corrections to be at most of the same order of magnitude of the light neutrino masses.  This becomes the most important bound for $M_4>\ord{1\,{\rm GeV}}$. These bounds are compared with present constraints for direct searches and lepton flavour violating processes. For example, we find that for $M_4=10$~GeV the mass splitting must be below 1~MeV for active-heavy mixing close to the experimental limit.

\section{LNV at Electron-Positron Colliders}
\label{sec:hdecay}

In the previous section we have shown that strong limits to heavy neutrino mass differences and LNV exist. This points towards heavy neutrinos being pseudo-Dirac fermions in the region of the parameter space where this scenario can be tested in future colliders. It might seem hopeless in this situation to be able to determine the mass splitting  by kinematical methods. However, the Majorana nature of the heavy neutrinos is essential to establish their connection to the light neutrino masses. This motivates the exploration of other collider observables that are sensitive to LNV processes, which can either determine if the  heavy neutrinos are Majorana particles, and/or set stronger bounds on the heavy neutrino mass difference.

One such observable is the ratio of same-sign over opposite-sign dileptons~\cite{Anamiati:2016uxp,Antusch:2017ebe,Das:2017hmg}, due to heavy neutrino oscillations. In this paper, we consider instead a possible asymmetry in the pseudorapidity distribution of a charged lepton coming from heavy neutrino decay. Specifically, we will focus on  $e^{-}e^{+}\to \nu N^{*} \to \nu\, \mu\, W^{(*)}$. In this section, we shall discuss this process in depth, without being concerned about the experimental signature nor the backgrounds, which will be taken into account in the following section. 

\begin{figure}[tb]
\centering
\includegraphics[width=0.35\textwidth]{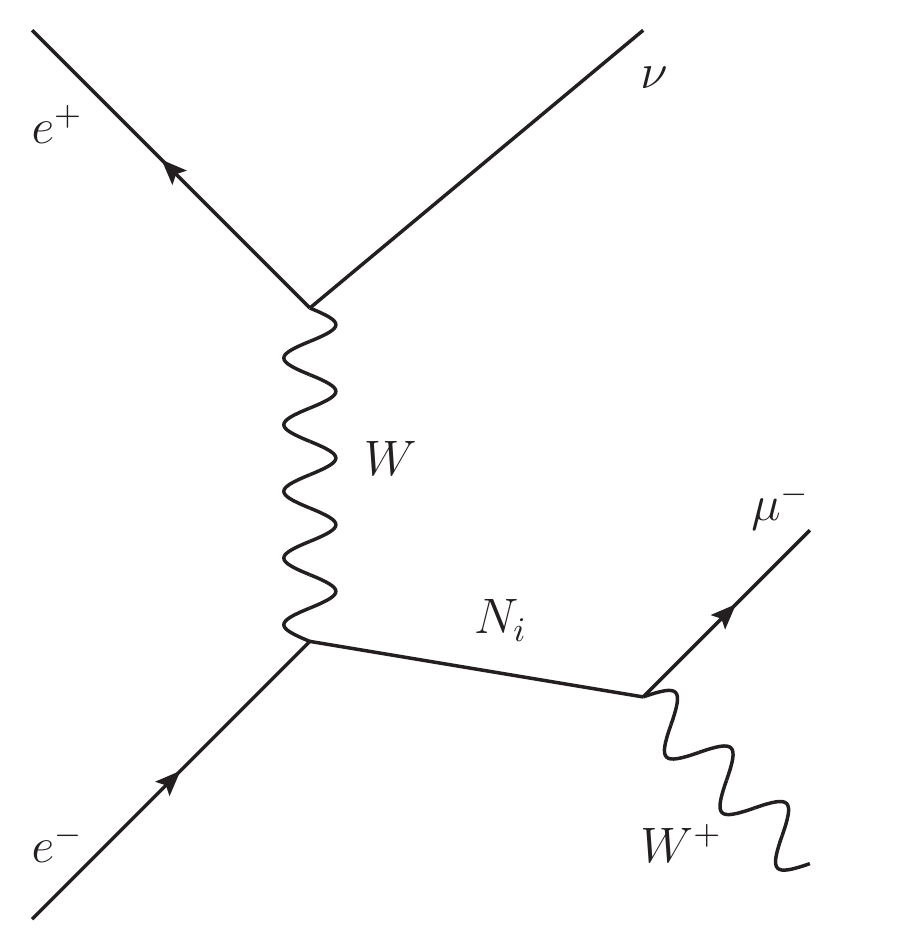} \hspace{1cm}
\includegraphics[width=0.35\textwidth]{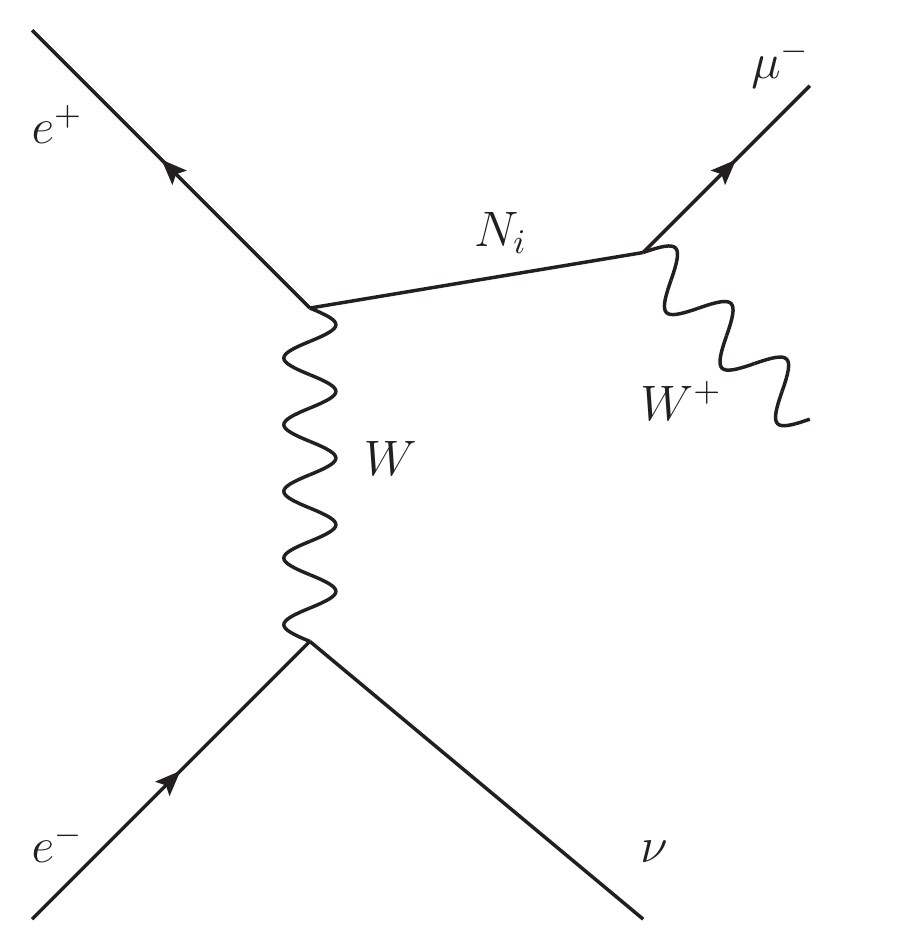}
\caption{\label{fig:diagrams} The process $e^{-}e^{+}\to \nu N^{*} \to \nu\, \mu\, W^{(*)}$. Diagram A (left) conserves lepton number, Diagram B (right) does not. }
\end{figure}

Let us consider the production of a negatively charged muon. The process can be mediated by different amplitudes, such as those shown in Fig.~\ref{fig:diagrams}.  The diagram on the left in Fig.~\ref{fig:diagrams}  is lepton number conserving (LNC), and it occurs regardless of $N$ being Dirac or Majorana. In contrast, the diagram on the right is LNV, and can only occur if $N$ is Majorana. Both contributions cannot be told apart because of the presence of a light left- or right-handed neutrino in the final state, which is unobserved. However, we shall see in the next section that the pseudorapidity distribution of the lepton will be different~\cite{delAguila:2005pin} and therefore this observable is a discriminator of the LNV and LNC situations.

We consider the two contributions to the amplitude for the process in Fig.~\ref{fig:diagrams} and will show explicitely how the B diagram vanishes for pseudo-Dirac neutrinos in the LNC limit.  Their amplitudes are~\cite{Denner:1992me}:
\begin{eqnarray}
\mathcal{M}_{A}&=&\left(\frac{g}{\sqrt{2}}\right)^3\sum_{j=4}^5 U_{ej}U^{*}_{\mu j}U^{*}_{e\nu}\left[\bar{u}_{\mu}(p_3) \gamma^{\lambda}P_L\, S_j\, \gamma^{\mu}P_L\, u_e(p_1)\right]\left[\bar{v}_{\bar e}(p_2)\gamma^{\nu} P_L\, v_\nu(p_5)\right]D_{\mu\nu}(p_A)\,\epsilon^{*}_{\lambda}(p_4), \nonumber \\ \\
\mathcal{M}_{B}&=&-\left(\frac{g}{\sqrt{2}}\right)^3\sum_{j=4}^5 U^*_{e j}U^{*}_{\mu j}U_{e\nu} \left[\bar{v}_{\bar e}(p_2)\gamma^{\nu} P_L\, S_j\, \gamma^{\lambda} P_R\,v_{\mu}(p_3)\right]\left[\bar{u}_\nu(p_5)\gamma^{\mu} P_L\,u_e(p_1)\right]D_{\mu\nu}(p_B)\,\epsilon^{*}_{\lambda}(p_4), \nonumber \\
\end{eqnarray}
where we call  $p_1,\,\ldots,\,p_5$ the momenta of $e^-$, $e^+$, $\mu^-$, $W^{+(*)}$, $\nu$  respectively.
Here, the term $\epsilon^*_{\lambda}(p_4)$ represents the final state coming from the $W^{(*)}$. If the $W$ is on-shell, it would be a polarization vector, otherwise it represents an additional $W$ propagator coupled to a fermion current. The propagator of each virtual heavy neutrino $N_j$, with mass $M_j$ and width $\Gamma_j$ is:
\begin{equation}
-i S_j=\frac{\slashed{q}+M_j}{q^2-M^2_j+iM_j\,\Gamma_j} \equiv\frac{\slashed{q}+M_j}{f(M_j)},
\end{equation}
with $q=p_3+p_4$. In our calculation, we have written the virtual $W$ propagator $D_{\mu\nu}$ in the unitary gauge, which can depend on $p_A=p_5-p_2$ or $p_B=p_5-p_1$.

Direct inspection shows that  the interference terms between A and B amplitudes  are proportional to the masses of the light neutrinos, so they can be safely neglected. The total unpolarized amplitude squared is therefore of the form
\begin{equation}
\vert \mathcal{M} \vert ^2=|\mathcal{M}_A|^2+ |\mathcal{M}_{B}|^2~,
\end{equation}
with
\begin{eqnarray}
\label{eq:M1}
|\mathcal{M}_{A}|^2&=& \frac{1}{4}\left(\frac{g}{\sqrt{2}}\right)^6 \left[\sum_{j,k=4}^5\Omega_{Aj}\Omega^*_{Ak}\right] G^{\lambda\delta}_{A}\, \epsilon^{*}_{\lambda}(p_4)\,\epsilon_{\delta}(p_4), \\
\label{eq:M4}
|\mathcal{M}_{B}|^2 &=&\frac{1}{4}\left(\frac{g}{\sqrt{2}}\right)^6 \left[\sum_{j,k=4}^5\frac{M_jM_k}{q^2}\Omega_{Bj}\Omega_{Bk}^*\right] G^{\lambda\delta}_{B}\,\epsilon^{*}_{\lambda}(p_4)\,\epsilon_{\delta}(p_4).
\end{eqnarray}
Here, we have defined:
\begin{eqnarray}
G^{\lambda\delta}_{A} &\equiv& {\rm Tr}[ \gamma^{\lambda}\slashed{q}\gamma^{\mu}P_L\, \slashed{p}_{1}\gamma^{\beta}\slashed{q}\gamma^{\delta}\slashed{p}_{3}]\,{\rm Tr}[\gamma^{\nu}\slashed{p}_{5}P_R\gamma^{\alpha}\slashed{p}_{2}]D_{\mu\nu}(p_A)D_{\alpha\beta}(p_A) \\
G^{\lambda\delta}_{B} &\equiv& q^2\,{\rm Tr}[ \gamma^{\nu}\gamma^{\lambda}\slashed{p}_{3}P_L\, \gamma^{\delta}\gamma^{\alpha}\slashed{p}_{2}]{\rm Tr}[\gamma^{\mu}\slashed{p}_{1}P_R \gamma^{\beta}\slashed{p}_{5}] D_{\mu\nu}(p_B)D_{\alpha\beta}(p_B)
\end{eqnarray}
and:
\begin{align}
\Omega_{Aj}\equiv\frac{U^{*}_{\mu j}U_{e j}U^{*}_{e \nu}}{f(M_j)}~, & &
\Omega_{Bj}\equiv\frac{U^{*}_{\mu j}U^{*}_{e j}U_{e \nu}}{f(M_j)}~.
\end{align}

The Majorana nature of the heavy neutrino is revealed by the presence of the B contribution. Thus, we expect $|\mathcal{M}_{B}|^2$ to vanish in the LNC limit, in which the mass splitting goes to zero:
\begin{eqnarray}
M_5 \rightarrow M_4,\;\;\; \Gamma_5 \rightarrow \Gamma_4. 
\label{eq:lncl}
\end{eqnarray}

Let us first analyse the A contribution, which is proportional to the term:
\begin{equation}
\Phi_A\equiv \sum_{jk}\Omega_{Aj}\Omega^*_{Ak} = |U_{e \nu}|^2 \sum^{5}_{k,j=4} \frac{U^{*}_{\mu j}\,U_{e j}\,U_{\mu k}\,U^{*}_{e k} }{f(M_j)f^*(M_k)}.
\end{equation}
Writing the mixings in our parametrization (see Eqs.~(\ref{eq:mixingmats}) and (\ref{eq:mixapprox})) we obtain:
\begin{equation}
\Phi_A = m^{2}_{3}\,|Z_{e}|^2|Z_{\mu}|^2|U_{e \nu}|^2\cosh^{4}\gamma_{45}
\left(\frac{1}{M^{2}_{4}|f(M_{4})|^2} +
 \frac{2}{M_{4}M_{5}}\Re e\left[\frac{1}{f(M_{4})f^*(M_{5})}\right]+
 \frac{1}{M^{2}_{5}|f(M_{5})|^2}\right),
 \label{eq:phia}
\end{equation}
where the interference of the contribution from the two virtual neutrinos is explicitely shown. 
 In the LNC limit,  Eq.~(\ref{eq:lncl}), we get the non-vanishing result:
\begin{equation}
\label{eq:phi1.limit}
\Phi_A\xrightarrow{LNC}  4|Z_{e}|^2|Z_{\mu}|^2|U_{e \nu}|^2 \cosh^{4}\gamma_{45}\,
\frac{m_3^2}{M_4^2}\frac{1}{|f(M_{4})|^2}.
\end{equation}

For the B contribution, we find instead:
\begin{eqnarray}
\Phi_B&\equiv& \sum_{jk}\frac{M_jM_k}{q^2}\Omega_{Bj}\Omega_{Bk}^* \nonumber \\
\label{eq:before-expansion.nice}
&=& m^{2}_{3}\,|Z_{e}|^2|Z_{\mu}|^2|U_{e \nu}|^2\cosh^{4}\gamma_{45}\left(\frac{1}{q^2|f(M_{4})|^2} -
\frac{2}{q^2}\Re e\left[\frac{1}{f(M_{4})f^{*}(M_{5})}\right]
 +\frac{1}{q^2|f(M_{5})|^2}   
\right),
\label{eq:phib}
\end{eqnarray}
which goes to zero in the LNC limit, as expected. 
 In order to properly understand the behaviour of $\Phi_B$ near this limit, we Taylor expand in $\delta M =M_5-M_4$, and $\delta \Gamma = \Gamma_5-\Gamma_4$. We find the first non-vanishing term at second order:
\begin{equation}
\Phi_{B}  \xrightarrow{LNC}
4|Z_{\mu}|^2|Z_{e}|^2|U_{e \nu}|^{2}\cosh^{4}\gamma_{45}\,\frac{m_3^2}{M_4^2}\frac{M_4^4}{q^2|f(M_4)|^4}\left[
\left(1+\frac{\Gamma_4^2}{4M_4^2}\right)(\delta M)^{2}  +
\frac{1}{4}(\delta \Gamma)^{2}+
\frac{\Gamma_4}{2M_4}\delta \Gamma\,\delta M\right].
\end{equation}

We can set the heavy neutrino to be on-shell by taking $q^2\to M_4^2$, which implies that $|f(M_4)|^2\to M_4^2\,\Gamma_4^2$. In this limit, we can compare both contributions in a straightforward way:
\begin{equation}
\left(\frac{\Phi_B^{}}{\Phi_A}\right)_{\rm on-shell}\xrightarrow{LNC}
\left(1+\frac{\Gamma_4^2}{4M_4^2}\right)\left(\frac{\delta M}{\Gamma_4}\right)^2  +
\frac{1}{4}\left(\frac{\delta \Gamma}{\Gamma_4}\right)^2+
\frac{\Gamma_4}{2M_4}\frac{\delta\Gamma\,\delta M}{\Gamma_4^2}.
\end{equation}
We find that, for $\Phi_B$ not to be negligible in front of $\Phi_A$ in the LNC limit, one needs at least one of the ratios $\delta M/\Gamma_4$ or $\delta \Gamma/\Gamma_4$ to be non-vanishing. In practice $\delta M \gg \delta \Gamma$ and therefore the ratio is controlled by $(\delta M/\Gamma_4)^2$. This result is to be expected since the cancellation of the LNV contribution requires the interference of the amplitudes mediated by the two heavy neutrino states. Such interference can only occur if $\delta M$ is sufficiently smaller than the decay width $\Gamma_4$. 

In the opposite limit, when $\delta M \gg \Gamma_4$, the interference terms in Eqs.~(\ref{eq:phia}) and (\ref{eq:phib}) are strongly suppressed due to the negligible overlap of the two Breit-Wigners peaked at $q^2 = M_4^2$ and $M_5^2$. Only the first and third terms contribute when $q^2=M^2_4$ or $q^2= M_5^2$ respectively and  $\Phi_B \simeq \Phi_A$ in this case.

We note that the total rate is no different in the two limiting cases, since $\Phi_A^{LNV} \simeq \Phi_B^{LNV}$, while $\Phi_A^{LNC} \simeq 2 \Phi_A^{LNV}$,  and $\Phi_B^{LNC} \simeq 0$.

\section{Forward-Backward Asymmetry at the ILC}
\label{sec:ILC}

\begin{figure}
	\centering
\includegraphics[width=0.48\textwidth]{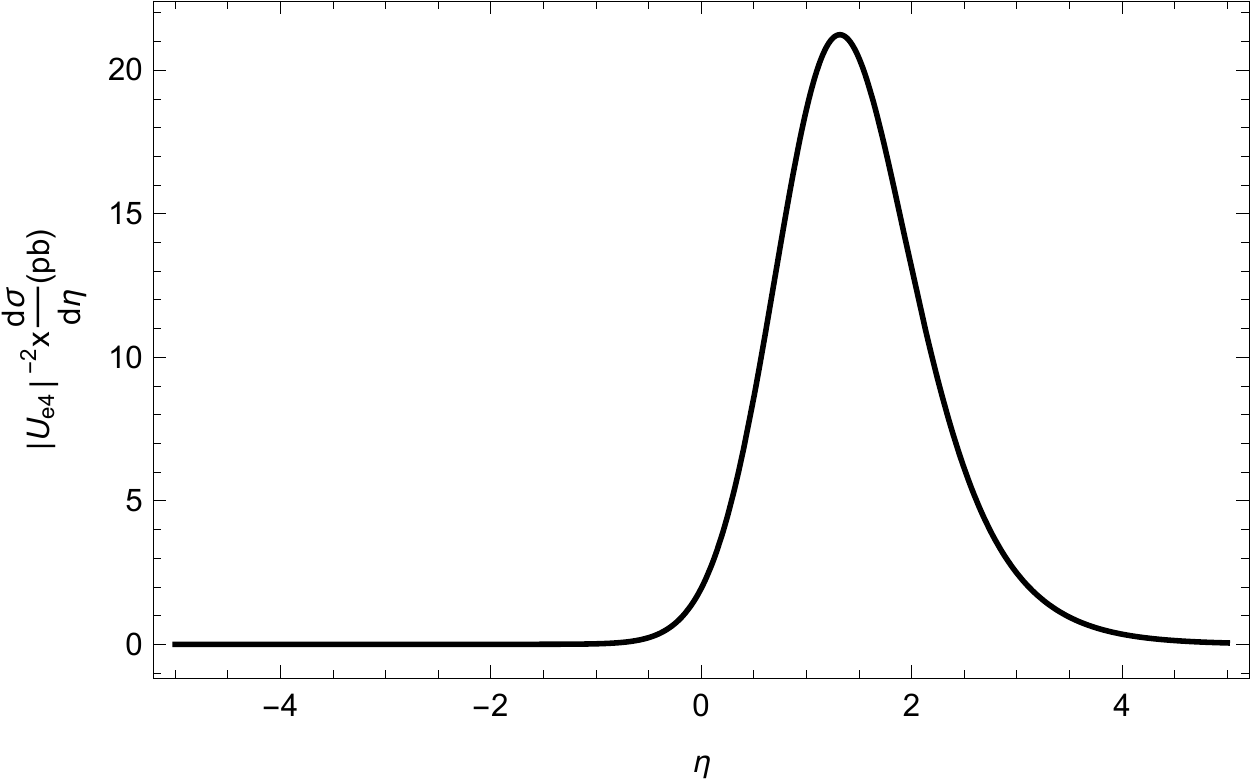} \hfill
\includegraphics[width=0.48\textwidth]{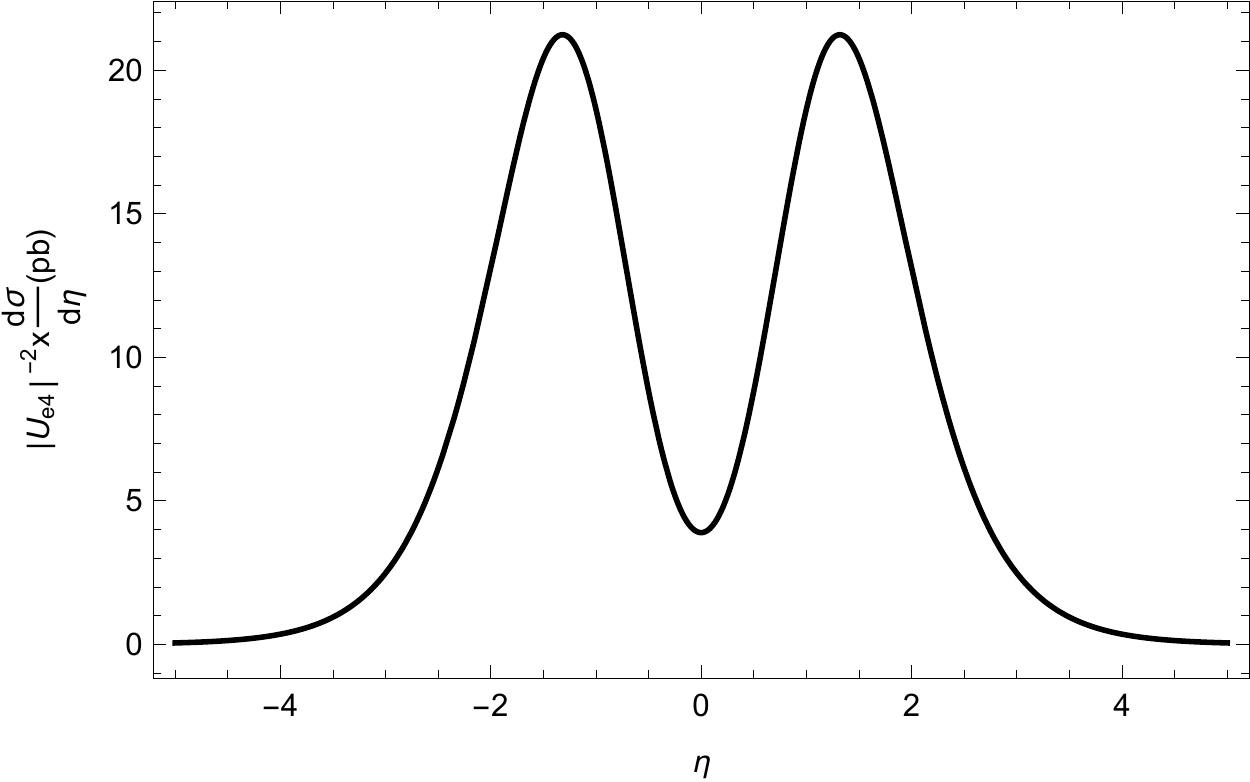}	
\caption{Pseudo-rapidity distribution of $N_i$ in the W-exchange process involving a $\bar{\nu}$ (left) and both a $\bar\nu$ and a $\nu$ (right).}
\label{fig:dsde}
\end{figure}

We now consider the pseudorapidity distribution of the final lepton, $\ell = e$ or $\mu$ for each charge separately. To understand the origin of the asymmetry in this distribution,  we can start by considering the pseudorapidity distribution of the heavy pseudo-Dirac neutrino in the two diagrams of Fig.~\ref{fig:diagrams}, i.e.\ in combination with  a light neutrino, which can be either left-handed $\nu$ or right-handed ${\bar \nu}$.
Any asymmetry in the pseudorapidity of the heavy neutrino will be inherited by the final lepton due to the boost. The contribution of $W$ exchange to the unpolarized differential cross section  for the process $e^+ e^- \rightarrow {\bar \nu} N_i$ (neglecting the electron and light neutrino masses) is given by:
\begin{eqnarray}
{d \sigma\over d \cos\theta} = { s-M_i^2\over 32 \pi s^2} \langle |{\mathcal M}|^2 \rangle, 
\end{eqnarray}
with $\theta$ the angle between the heavy neutrino and the incoming electron. The amplitude squared is:
\begin{eqnarray}
\langle |{\mathcal M}|^2 \rangle = \left({g\over \sqrt{2}}\right)^4 |U_{e4}|^2 {(s+t) (s+t -M_i^2)\over (t-M_W^2)^2},
\end{eqnarray}
where $s, t$ are the Mandelstam variables. Changing variables to the pseudorapidity of the heavy neutrino:
\begin{eqnarray}
\eta = - \ln\left(\tan {\theta\over 2}\right), \;\;\; \cos\theta= \tanh \eta,
\end{eqnarray}
we find 
\begin{eqnarray}
{d \sigma \over d \eta} = {1 \over (\cosh\eta)^2} {d \sigma \over d\cos\theta}, 
\end{eqnarray}
which is shown of the left panel of Fig.~\ref{fig:dsde} for $\sqrt{s} = 250$~GeV and $M_i= 5$~GeV. The asymmetry varies very little with the mass of the $M_i$ but is very sensitive to $\sqrt{s}$. 

If we define the pseudorapidity  asymmetry as
\begin{eqnarray}
A_\eta \equiv {\int_0^\infty d\eta \frac{d\sigma}{d\eta}-\int_{-\infty}^0 d\eta {d\sigma\over d\eta}\over \int_0^\infty d\eta {d\sigma\over d\eta}+\int_{-\infty}^0 d\eta {d\sigma\over d\eta}},
\end{eqnarray}
we get  $A_\eta= 0.9743$.
The subleading $Z$-exchange contribution at this center of mass energy, $e^+ e^- \to Z^* \to{\bar\nu}\,N_i$  also gives an asymmetry, but it is smaller. 

The contribution of the process $e^+ e^-\to\nu N_i$ via $W$ exchange gives exactly the opposite distribution in pseudorapidity, so that the sum of the two contributions  gives the result on the right panel of Fig.~\ref{fig:dsde}, that is, zero asymmetry.

Since the $N_i$ has a significant boost, the decay products follow the same angular distribution. We expect therefore a significant asymmetry in pseudorapidity of leptons with a given charge in the Dirac case, where only one diagram contributes, versus the Majorana case where both do equally. 

The heavy neutrino production is expected to have a very large background coming from the SM process $e^-e^+\to W^+W^-$. In order to avoid this background, we require the heavy neutrino to be nearly on-shell, with a large enough lifetime in order to decay far from the interaction point. Experimentally, this leaves a displaced vertex signature, which has been studied extensively in the literature~\cite{Helo:2013esa,Blondel:2014bra,Cui:2014twa,Gago:2015vma,Duarte:2016caz,Antusch:2016vyf,Caputo:2017pit,Antusch:2017hhu,Abada:2018sfh}. To be able to observe this signature at the LHC or future colliders, the lifetime needs to be large enough, which requires heavy neutrinos with masses between $1-50$~GeV.

In order  to study the prospects of measuring this lepton asymmetry at the ILC, we have implemented  the model in \texttt{SARAH~4.13.0}~\cite{Staub:2008uz,Staub:2012pb,Staub:2013tta}, with the calculation of the mass spectrum and decay widths carried out in \texttt{SPheno~4.0.3}~\cite{Porod:2003um,Porod:2011nf}. The output of both programs was input into \texttt{WHIZARD~2.6.0}~\cite{Kilian:2007gr,Moretti:2001zz}, which generated $e^+e^-$ interactions at the ILC. The simulation included a polarization of $(0.80,\,0.30)$ for the initial state electrons and positrons, respectively, as well as ISR and beamstrahlung. Following the reports in~\cite{Fujii:2017vwa,Asai:2017pwp}, the collisions were produced at a center-of-mass energy of 250~GeV, with a final integrated luminosity of 2~ab$^{-1}$. As our final state involves quarks, coming from $W^*(p_4)$, the parton shower and hadronization of the jets was carried out with the built-in version of \texttt{Pythia~6}~\cite{Sjostrand:2006za}. For the detector simulation and reconstruction of events, we used \texttt{DELPHES~3.4.1}~\cite{deFavereau:2013fsa,Cacciari:2011ma}, with the DSiD card~\cite{Potter:2016pgp}. 
 

\begin{table}
\begin{center}
\begin{tabular}{|c| c | c | c | c | c | c |}
\hline
Name & Mass (GeV) & $|U_{e4}|^2$ & $|U_{\mu4}|^2$ & $|U_{\tau4}|^2$ & $\Gamma_4$~(meV) & $c\,\tau_4$~(mm) \\
\hline
\textit{Light} & 5 & $1.5\times10^{-6}$ & $1.0\times10^{-5}$ & $1.3\times10^{-5}$ & $0.02$ & 10 \\
\hline
\textit{Heavy} & 20 & $7.5\times10^{-7}$ & $5.0\times10^{-6}$ & $6.3\times10^{-6}$ & $20$ & 0.01 \\
\hline
\end{tabular}
\end{center}
\caption{\label{Table:Benchmarks} Benchmark scenarios considered in our study for a normal light neutrino hierarchy. We show masses, mixing, decay width and decay length.}
\end{table}

The procedure for establishing the cuts on the displaced vertex follows the discussion in~\cite{Antusch:2016vyf,Accomando:2016rpc,Caputo:2017pit}, with the heavy neutrino momentum being reconstructed from the  parton-level quark and charged lepton momenta. With this, as well as with the heavy neutrino lifetime, we use the appropriate probability distribution to randomly assign a position for the secondary vertex. This position must be contained within the detector, that is, if $L_T$ and $L_z$ are, respectively, the transverse and longitudinal coordinates of this vertex, we require:
\begin{align}
 L_T< 2.49~{\rm m}, \;\;\;\;\;\; &  L_z < 3.018~{\rm m}.
\end{align}
In addition, we need $L_T> 10~{\mu\rm m}$, in order to avoid SM backgrounds from long-lived meson decays~\cite{Antusch:2016vyf}. However, the most important cut is on the impact parameter $d_\ell$ of the charged lepton on the final state, given by:
\begin{equation}
 d_{\ell}\equiv\frac{L_x\,p^\ell_y-L_y\,p^\ell_x}{p^\ell_T} > 6~{\mu\rm m}
\end{equation}
where $L_{x,y}$ and $p^\ell_{x,y}$ are the components of $L_T$ and $p_T^{\ell}$, respectively, on the X and Y axes.

In what follows, we consider two benchmark scenarios (\textit{light} and \textit{heavy}), which have a high probability of satisfying the previous constraints. These benchmarks differ by the heavy neutrino masses and mixing, with further details given in Table~\ref{Table:Benchmarks}.

\begin{figure}[tb]
\centering
\includegraphics[width=0.45\textwidth]{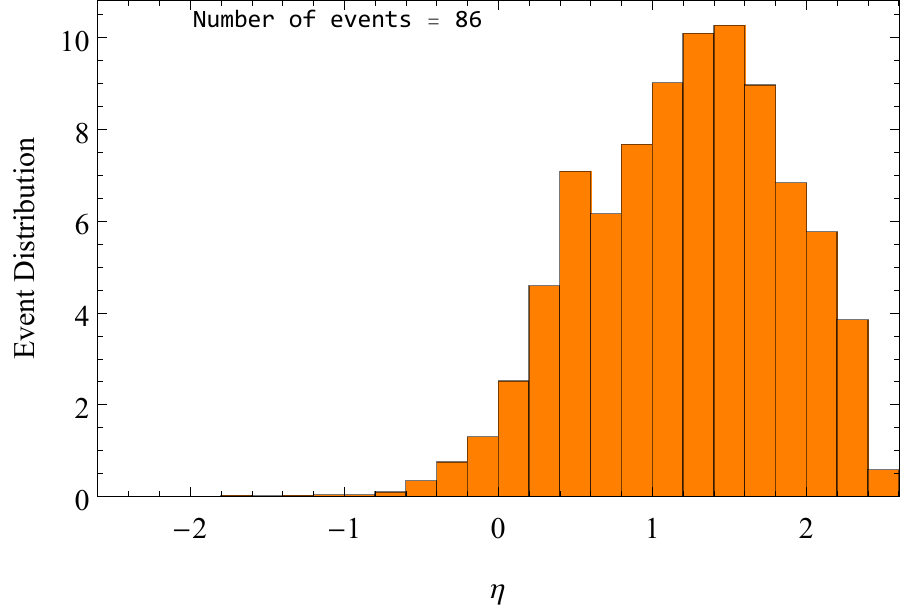}\hfill
\includegraphics[width=0.45\textwidth]{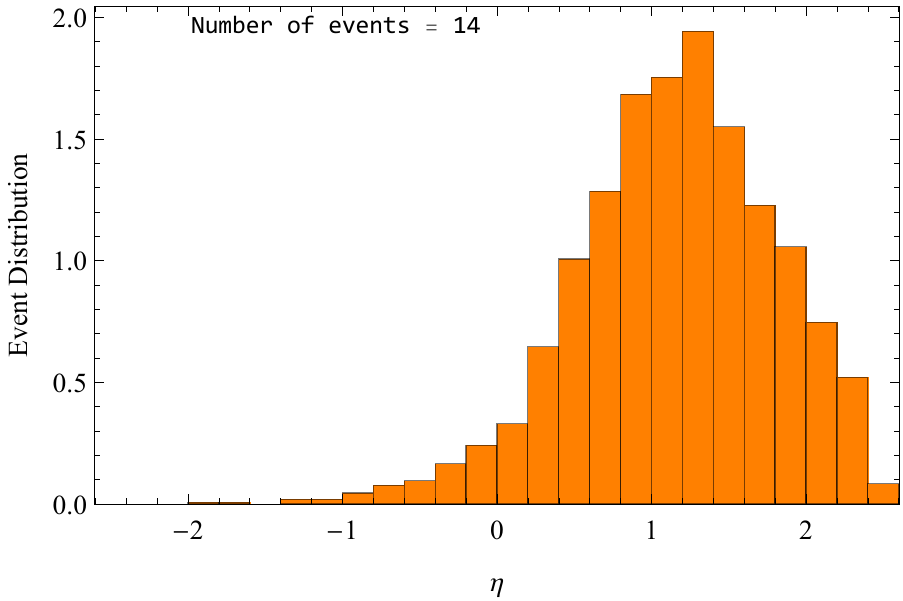} \\
\includegraphics[width=0.45\textwidth]{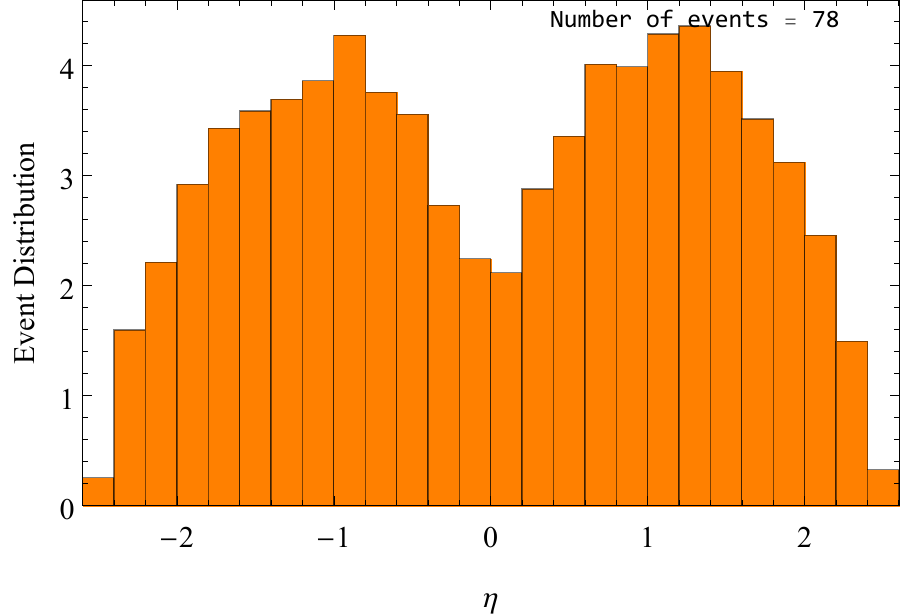}\hfill
\includegraphics[width=0.45\textwidth]{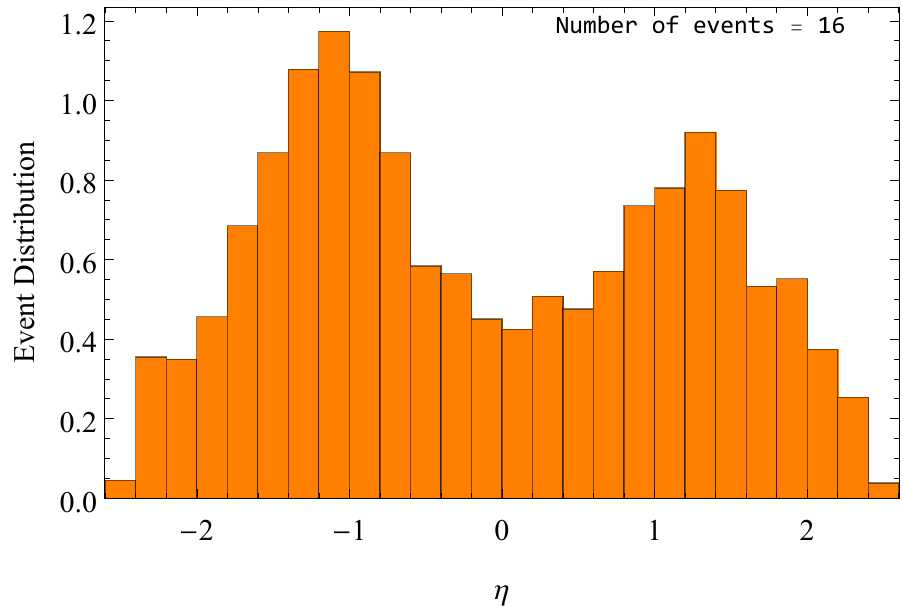}
\caption{Pseudorapidity distribution of the charged leptons in $e^+e^-\to\nu\,N^*\to\nu\,\ell+{\rm jets}$, for the \textit{light} benchmark on the left column, and the \textit{heavy} benchmark on the right column. We show $\delta M\ll\Gamma_4$ and $\delta M\gg\Gamma_4$ on the upper and lower rows. Each panel shows the total number of $\ell^-$ events, we have a similar number of $\ell^+$ events.} 
\label{fig:histo}
\end{figure}
In Figure~\ref{fig:histo}, we present the pseudorapidity distribution of the $\ell^-$, for both benchmarks. We show results for different mass splittings, depending on the value of $\Gamma_4$. The top row shows the distribution for $\delta M\ll\Gamma_4$. Following our reasoning from Section~\ref{sec:hdecay}, we expect the LNV contribution for this process to be negligible, such that the $\ell^-$ will be necessarily produced in association with a ${\bar \nu}$. As a consequence, the $\ell^-$ should follow the pseudorapidity distribution shown on the left panel of Figure~\ref{fig:dsde}, which is what we observe. Similarly, the $\ell^+$ will be produced in association with a $\nu$, and follows the opposite distribution. This means that, in this situation, two opposite forward-backward asymmetries can be expected, one for the $\ell^-$ and one for the $\ell^+$.

The bottom row shows the same pseudorapidity distribution when $\delta M\gg\Gamma_4$. Here the LNV contribution is larger, such that the $\ell^-$ can be produced in association with either a $\nu$ or a ${\bar \nu}$, equally favouring both signs of pseudorapidity. As the same behaviour is observed for $\ell^+$, this leads to the vanishing of both asymmetries. This confirms that the asymmetry depends directly on the mass difference $\delta M$, such that the latter can be constrained by the observation of the former.

\begin{figure}[tbp]
\centering
\includegraphics[width=0.48\textwidth]{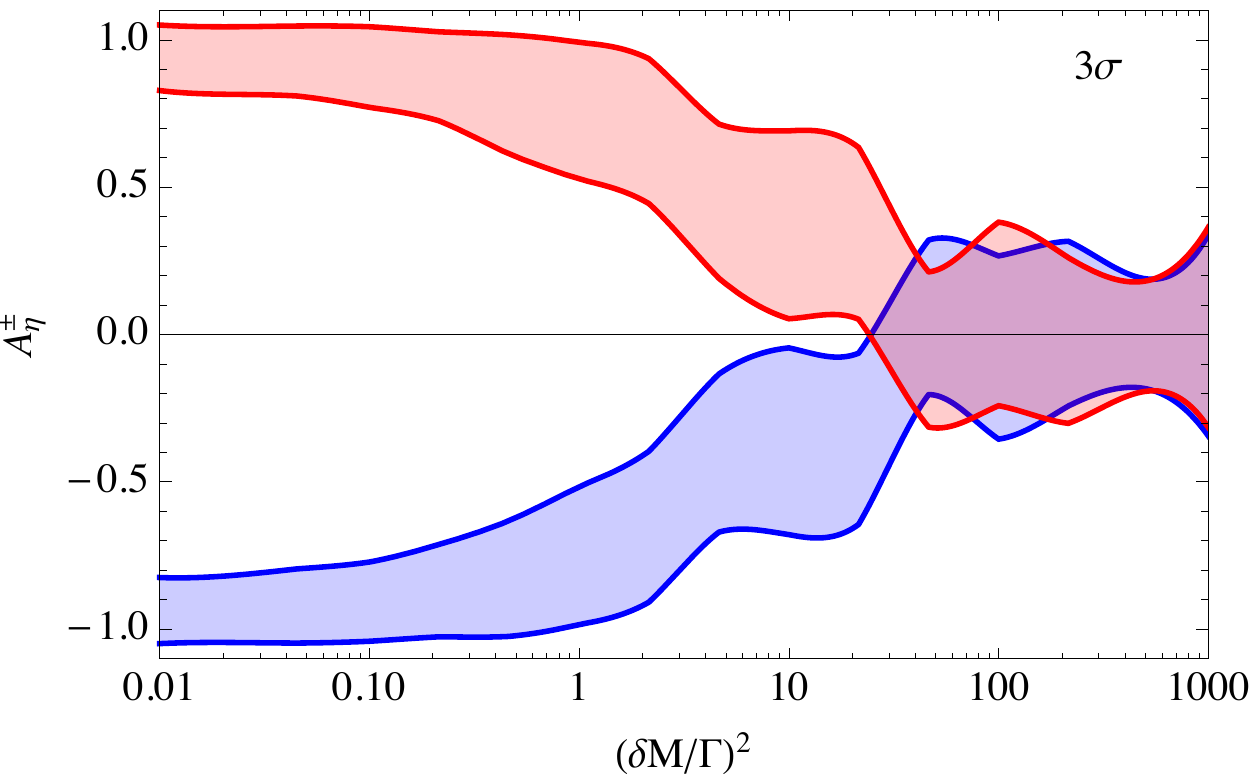}\hfill
\includegraphics[width=0.48\textwidth]{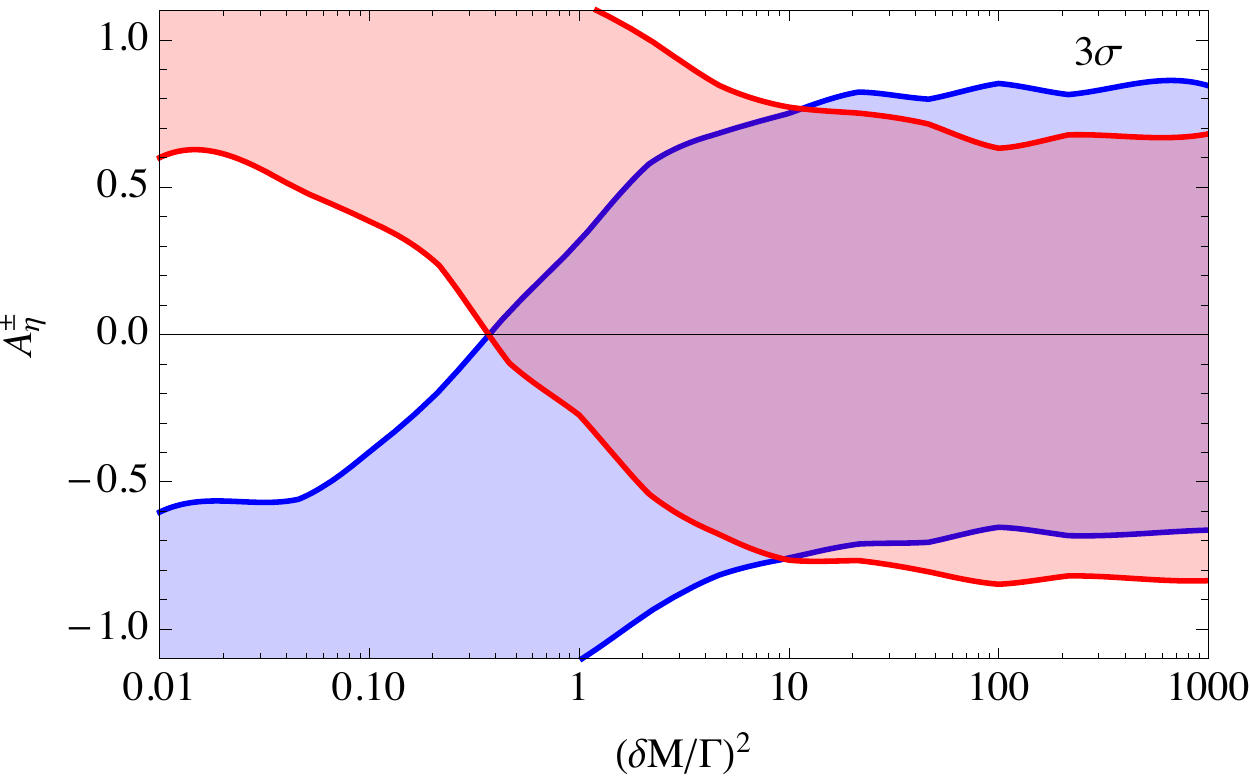}
\caption{Forward-backward asymmetry as a function of $(\delta M/\Gamma)^2$. We show the $3\sigma$ region for $A_\eta^-$ ($A_\eta^+$) in red (blue). The \textit{light} and \textit{heavy} benchmarks are shown on the left and right, respectively.} 
\label{fig:assym}
\end{figure}
In order to quantify this statement, we define the forward-backward asymmetry $A_\eta^\pm$ for a lepton with specific charge as:
\begin{equation}
 A^{\pm}_\eta=\frac{N^{\pm}(\eta>0)-N^{\pm}(\eta<0)}{N^\pm_{\rm tot}}~,
\end{equation}
where $N^\pm(\eta>0)$ and $N^\pm(\eta<0)$ are the number of events where $\ell^\pm$ has positive or negative pseudorapidity, respectively, and $N^\pm_{\rm tot}=N^\pm(\eta>0)+N^\pm(\eta<0)$.

In Figure~\ref{fig:assym}, we calculate $A_\eta^\pm$ for several values of $(\delta M/\Gamma_4)^2$ and interpolate the results. The shaded regions indicate the $3\sigma$ confidence intervals, evaluated by taking into account the expected number of events. We find that the behaviour on the LNV and LNC limits matches our expectations, that is, $A^\pm\to0$ when $(\delta M/\Gamma_4)^2\to\infty$, and $|A^\pm|\sim1$ for $(\delta M/\Gamma_4)^2\to0$.

For the \textit{light} benchmark, we have a relatively large enough number of events, so the asymmetry can be  determined with good precision. At 3$\sigma$,  $A_\eta^\pm$ is compatible with zero for $(\delta M/\Gamma_4)^2\gtrsim20$, and  with $\pm1$ for $(\delta M/\Gamma_4)^2\lesssim1$. Therefore, (not) observing the asymmetry establishes upper (lower) limits on $\delta M$, depending on $\Gamma_4$. In addition, we have a region where $|A^\pm_\eta|$ might be measured to be neither zero nor unity. In this case, the splitting could be 
constrained as $\ord{1}<(\delta M/\Gamma_4)^2<\ord{10}$. Such an observation would be particularly interesting in connection to resonant leptogenesis models.

On the other hand, the \textit{heavy} benchmark has much less events and the precision is poorer. The asymmetry is compatible with zero for $(\delta M/\Gamma_4)^2\gtrsim0.3$ and with unity when $(\delta M/\Gamma_4)^2\lesssim1$. Here we can again place upper or lower bounds on $\delta M$, provided we know $\Gamma_4$.  In this case there is not enough precision to measure $|A_\eta^\pm|$ to be different from both zero and unity at $3\sigma$. 

We now proceed to quantify the  hypothetical bound on $\delta M$, based on the observation, or not, of a pseudorapidity asymmetry by combining the data for the two charges:
\begin{equation}
 A^{\rm tot}_\eta=\frac{A^-_\eta-A^+_\eta}{2},
\end{equation}

For the \textit{light} benchmark, still quoting $3\sigma$ errors, we find $A_\eta^{\rm tot}=0.94\pm0.09$ when $(\delta M/\Gamma_4)^2=10^{-2}$, $A_\eta^{\rm tot}=0.75\pm0.15$ for $(\delta M/\Gamma_4)^2=1$, and $A_\eta^{\rm tot}=0.01\pm0.24$ if $(\delta M/\Gamma_4)^2=10^3$. Observing the asymmetry at $3\sigma$ would mean that $(\delta M/\Gamma_4)^2\lesssim0.16$, which implies $\delta M\lesssim8$~$\mu$eV. In contrast, a symmetric distribution would be observed for $(\delta M/\Gamma_4)^2\gtrsim45$, which leads to $\delta M\gtrsim100$~$\mu$eV\footnote{Even though $\delta M$ has a strict lower bound due to the contribution of light neutrino masses, this is significantly smaller than the limits we are obtaining for these test points.}.

Similarly, for the \textit{heavy} benchmark, we have $A_\eta^{\rm tot}=0.92\pm0.08$ if $(\delta M/\Gamma_4)^2=10^{-2}$, $A_\eta^{\rm tot}=0.41\pm0.17$ for $(\delta M/\Gamma_4)^2=1$, and $A_\eta^{\rm tot}=-0.08\pm0.18$ when $(\delta M/\Gamma_4)^2=10^3$. The observation of $A_\eta^{\rm tot}$ compatible with unity gives $(\delta M/\Gamma_4)^2\lesssim0.53$, or $\delta M\lesssim15$~meV. Moreover, an observation compatible with zero implies $(\delta M/\Gamma_4)^2\gtrsim0.72$, meaning that $\delta M\gtrsim17$~meV.

It is clear that these hypothetical constraints, of the order of $\mu$eV (meV) for the \textit{light} (\textit{heavy}) benchmark, are significantly stronger than any of those obtained in Section~\ref{sec:constraints}. This test thus implies a powerful way to indirectly extract information on $\delta M$, and with it probe the nature of the heavy neutrinos as well as the breaking of LNV in Nature.

\section{Conclusions}

The minimal Type I seesaw adds two heavy neutrinos to the SM, providing mass to two light neutrinos. By choosing the model parameters appropriately, one can have relatively large active-heavy mixing even if the heavy neutrinos have masses of the order of the GeV. This is possible due to specific textures within the full neutrino mass matrix, motivated by the small breaking of Lepton Number, with LNV elements linked to $\delta M=M_5-M_4$, the mass splitting of the heavy neutrinos. Currently, constraints coming from neutrinoless double beta decay and loop corrections require, in the parameter space accessible to colliders, values of $\delta M/M_4$ between $\ord{10^{-2}}$ and $\ord{10^{-6}}$.

In this work we have considered heavy neutrino production at the ILC, with the heavy neutrinos decaying into a charged lepton and jets. We have demonstrated that LNV channels vanish when $\delta M\to0$, consistent with expectations. In addition, we have shown that, in the pseudo-Dirac limit, the ratio between LNV and LNC contributions is proportional to $(\delta M/\Gamma_4)^2$.

The presence of the LNV diagrams can be tested by observing the pseudorapidity distribution of the charged leptons into which the heavy neutrinos decay. We have shown that, when the LNV contribution is absent, the distribution of a lepton with specific charge is asymmetric. In contrast, when $\delta M$ is large and LNV cannot be neglected, the asymmetry is absent. Thus, the presence of an asymmetric pseudorapidity distribution can set bounds on  $\delta M$ as a function of $\Gamma_4$.

The process we explored in this work has very large backgrounds. To reduce them, we are forced to demand the existence of a displaced vertex. At the ILC, this requires values of $M_4$ between 1 and 30~GeV, with $\Gamma_4$ ranging from $\mu$eV to meV. This means that the values of $\delta M$ that we can probe are around this order. For $M_4 = 5$~GeV, the critical region where the asymmetry disappears is around $\delta M/M_4\sim\ord{10^{-14}}$. For $M_4 = 20$~GeV, the region is around $\delta M/M_4 \sim\ord{10^{-12}}$. Thus, the study of the asymmetry can determine if the mass splitting is smaller or larger than these scales, establishing bounds which are much more precise than those currently available.

\section{Acknowledgements}

P.H.~acknowledges support from grants FPA2017-85985-P, and  the European projects 
H2020-MSCA-ITN-2015//674896-ELUSIVES and H2020-MSCA-RISE-2015. J.J.P.~acknowledges funding by the {\it Direcci\'on de Gesti\'on de la Investigaci\'on} at PUCP, through grant DGI-2015-3-0026. O.S.N.~received funding from CienciActiva-CONCYTEC Grant 233-2015-1, as well as the grant {\it Becas J\'ovenes Investigadores 2017} from the \textit{Programa de Cooperaci\'on 07 de la Universitat de Val\`encia.}

\bibliographystyle{epjc}
\bibliography{pseudodirac}

\end{document}